\documentclass[12pt,a4]{article}

\usepackage{amsmath}

\usepackage{amsfonts}
\usepackage{amssymb}
\usepackage{graphicx}

\usepackage{amsmath,amssymb,graphicx,mathrsfs}
\usepackage[latin1]{inputenc}
\usepackage{bm}
\renewcommand{\H}{\mathcal{H}}
\newcommand{\be}{\begin{equation}}
\newcommand{\ee}{\end{equation}}
\newcommand{\bea}{\begin{eqnarray}}
\newcommand{\eea}{\end{eqnarray}}

\def\k{\boldsymbol{k}}

\def\cs2{c_{s}^{2}}

 \def\be{\begin{equation}}   \def\ee{\end{equation}}
 \def\ba{\begin{array}}      \def\ea{\end{array}}
 \def\bea{\begin{eqnarray}}   \def\eea{\end{eqnarray}}
 \def\bean {\begin{eqnarray*}}\def\eean{\end{eqnarray*}}

\begin{document}

\begin{flushright} {\footnotesize CERN-PH-TH/2011-076}  \end{flushright}
\vspace{5mm}
\vspace{0.5cm}
\begin{center}
\def\thefootnote{\fnsymbol{footnote}}

{\Large \bf Evolution Equation for  \\ Non-linear Cosmological Perturbations
}
\\[0.5cm]
{\large Ram Brustein$^{\rm a,b}$ and Antonio Riotto$^{\rm b,c}$}
\\[0.5cm]

\vspace{.2cm}

{\small\textit{$^{\rm a}$ Department of Physics, Ben-Gurion University, Beer-Sheva, 84105 Israel}}

\vspace{0.2cm}
{\small \textit{$^{\rm b}$  CERN, PH-TH Division, CH-1211,
Gen\`eve 23,  Switzerland}}

\vspace{.2cm}

{\small \textit{$^{\rm c}$  INFN Sezione di Padova, via Marzolo 8,
I-35131 Padova, Italy}}

\end{center}

\vspace{1.5cm}

\hrule \vspace{0.3cm}
{\small  \noindent \textbf{Abstract} \\[0.3cm]
We present a novel approach, based entirely on the gravitational potential, for studying the evolution of non-linear cosmological matter perturbations. Starting from the perturbed Einstein equations, we integrate out the non-relativistic degrees of freedom of the cosmic fluid and obtain a single closed equation for the gravitational potential. We then verify the validity of the new equation by comparing its approximate solutions to known results in the theory of non-linear cosmological perturbations.
First, we show explicitly that the perturbative solution of our equation matches the standard perturbative solutions. Next, using the mean field approximation to the equation, we show that its solution reproduces in a simple way the exponential suppression of the non-linear propagator on small scales due to the velocity dispersion.  Our approach can therefore reproduce the main features of the renormalized perturbation theory and (time)-renormalization  group approaches to the study of non-linear cosmological perturbations. We conclude by a preliminary discussion of the nature of the full solutions of the equation and their significance.

\vspace{0.5cm}  \hrule
\newpage
\def\thefootnote{\arabic{footnote}}
\setcounter{footnote}{0}

\section{Introduction}
\noindent
The  Large-Scale Structure (LSS) of the universe grows from an initial nearly scale-invariant spectrum of Gaussian fluctuations due to a gravitational instability.  The matter distribution of our universe today is well described on large scales by linear perturbation theory about a homogeneous and isotropic background. On length scales below about 10 Mpc the dynamics of matter is highly non-linear, so to describe its evolution in a qualitative way one has to resort numerical N-body simulations.

At  intermediate (quasi-linear) scales, the evolution of matter may be described analytically by extending standard perturbation theory (SPT) \cite{pr}.  In SPT one defines a series solution to the fluid equations in powers of the initial density field. The $n$-th order term of this series grows as the $n$-th power of the scale factor $a$ (for a pressureless fluid) which affects its convergence properties and limits its applicability.  Qualitative comparisons to simulations have shown that the domain of applicability of second-order  perturbation theory (in the linear power spectrum) is limited,  at redshift $z = 0$, to wavenumbers of about $0.05$ $h$/Mpc.

The need for improving theoretical predictions for the next generation of very large galaxy surveys has spurred many efforts to go beyond SPT. Through future observations of matter at intermediate scales, such as the observation of the  baryon acoustic oscillations at scales of the order of $10^2$ Mpc, one will be able to probe  the expansion history of the Universe and of the nature of dark energy. The statistical weight of galaxy surveys is higher at smaller scales, increasing quadratically with decreasing scales. Since the aim is to get to percent accuracy in observations and theoretical calculations the need of extending the domain of validity of the theory of non-linear cosmological perturbations to smaller scales is clear.

Renormalized perturbation theory (RPT) \cite{RPT} reorganizes the perturbation expansion in terms of different fundamental objects, the  so-called non-linear propagator and non-linear vertices, to improve the convergence of SPT. When the vertex is approximated by its lowest order (tree-level) form, then the matter power spectrum can be expressed as an expansion in terms of the non-linear propagator. The dominant contributions to the non-linear propagator are identified and summed explicitly in the large-momentum  limit, resulting in an exponential suppression of the power at short distance scales due to the matter velocity dispersion. This is the significant feature of RPT: the emergence of an intrinsic ultraviolet  cutoff which improves the convergence of the perturbation series at small distance scales.
Then, by matching the behavior of the non-linear propagator at high momentum with the one-loop propagator which is the a good approximation to the true propagator at low momenta one obtains a non-perturbative prediction for the non-linear propagator. Substituting this propagator in the first few diagrams of the reorganized expansion then gives a non-perturbative prediction for the power spectrum. The latter  is the sum of two terms: the first term is  exponentially suppressed and proportional to the initial spectrum and the second term represents the power generated by mode-coupling at smaller scales whose time- and scale-dependence are dictated by the non-linear propagator.

RPT  relies on the standard equations for the density and
velocity fields. An alternative approach is to use the fluid
equations to derive evolution equations for the power
spectrum and higher order correlators directly. For instance, the closure theory approach \cite{closure} approximates the three-point
correlator  by its leading order expression in
SPT. The Renormalization Group (RG) method \cite{MP} and its variant, in which time is the flow parameter \cite{MP1},
offers an alternative, and to some extent, a more compact way of getting the same results of RPT. Truncating the RG equations at the level of the four-point correlator leads to a solution that corresponds to the summation of an infinite class of perturbative corrections and reproduces the RPT results. The RG method can be also be extended to include higher-order correlators \cite{MPtri}, the effects of massive neutrinos \cite{MP2} and modified to account for non-Gaussian initial conditions \cite{MP3}.

Additional methods to extend SPT have been proposed. The path-integral formulation of the Vlasov equation  in terms of the distribution functions \cite{val}, in which statistical observables, such as the power spectrum, can be expressed as functional derivatives of a  generating functional. The Lagrangian resummation theory (LRT)  \cite{lag} reproduces the SPT power spectrum at the lowest non-trivial order and yields a non-perturbative prediction for the power spectrum that corresponds to resumming an infinite set of terms in SPT. The main difference between LRT and the RPT and RG methods is that in LRT the total power spectrum (including the mode-coupling piece) is exponentially suppressed at high momentum.

The  goal of this paper is to present a novel approach to study the non-linear evolution of matter perturbations which is entirely based  on the gravitational potential. The motivation for our approach is the fact that the dimensionless gravitational potential  is estimated to be smaller than about $10^{-5}$ on all cosmological scales of interest. The gravitational potential is of order unity only near black holes. For example, the gravitational potential in galaxies is a few times $10^{-7}$ and in clusters of galaxies it is of order a few times $10^{-5}$. Therefore, the gravitational potential is potentially a good  order parameter for a perturbative expansion. It is certainly a better expansion parameter than the matter density contrast which is of order unity already in the quasi-linear regime. Of course, linear perturbation theory for the gravitational potential does not suffice since space derivatives of the gravitational potential can be large and comparable to the background energy density. Here the fact that Einstein equations contain at most two space derivatives greatly restricts the number of terms that can appear in the full equation for the gravitational potential.

The basic tool of our approach is a non-perturbative equation for the gravitational potential which we obtain by  integrating  out the degrees of freedom of the non-relativistic fluid.  From the non-perturbative equation it is possible to reproduce in a few and simple steps the known SPT results. Furthermore, it is also possible to recover in a simple and physically transparent way the main results of the RPT and RG methods. In particular, the exponential suppression of the propagator in the high-momentum limit is recovered in our approach by solving our non-perturbative equation  in  the mean-field approximation. In this approximation the high-momentum modes of the perturbation ``see" the velocity fluctuations as a modification to background and the nature of the suppression acquires a clear physical interpretation.

Our equation provides a framework for evaluating the errors that one makes in various approximation schemes and allows to deduce some general conclusions about the evolution of non-linear cosmological perturbations even without a detailed analysis of its exact solutions. As we shall see, the  non-perturbative
equation  indicates that the gravitational potential (and therefore the density contrast)  can not grow indefinitely as time evolves.

The paper is organized as follows. In section 2 we derive the non-perturbative equation for the gravitational potential. In Section 3 we show how the SPT results, up to third-order, are reproduced from our equation. In Section 4 we discuss the mean-field solution of the non-perturbative equation, reserving Section 5 to the comparison with the RPT and RG methods. In section 6 we discuss in a qualitative way the nature of the full solutions of our equation and their possible significance. Finally, Section 7 contains our conclusions and expectations for additional directions into which our analysis can be extended.

\section{The non-perturbative equation for the gravitational potential}
\noindent
The goal of this section is to derive a non-perturbative equation for the gravitational potential in the presence of a fluid of non-relativistic dark matter particles. The extension
to the more realistic case of the cosmological constant (or dark energy) dominance is straightforward
and will be presented elsewhere.
As we have mentioned in the introduction, to derive a closed equation for  the gravitational potential we need to integrate out the fluid degrees of freedom, which we do using  Einstein equations.
Here we make use of the fact that the Einstein equations that the gravitational potential satisfies contain at most two space derivatives,
so if we wish to take them into account, it is enough to consider at most terms that are quadratic in the gravitational potential and contain two space derivatives. The rest of the terms are completely negligible on all relevant scales as explained in the introduction.

Our starting point is the metric in the Poisson gauge
\be
{\rm d} s^2=-a^2(\tau)\left[-(1+2\Phi)\,{\rm d}\tau^2+2\omega_i\,{\rm d}\tau{\rm d} x^i+\left((1-2\Psi)\delta_{ij}+\chi_{ij}\right)
{\rm d}x^i{\rm d}x^j\right]\, ,
\ee
where $a(\tau)$ is the scale factor as a function of the conformal time $\tau$, $\Phi$ and $\Psi$ are the
lapse function and the gravitational potential respectively, $\omega_i$ are the vector perturbations
and $\chi_{ij}$ the tensor ones. As we are interested in subhorizon scales,  we will retain  first-order terms in $\Psi$ and $\Phi$ and only the would-be second-order terms with  two space derivatives. In particular, the difference $(\Psi-\Phi) $ is  second-order in $\Psi$,  so we need to keep  $\nabla^2 (\Psi-\Phi)$.

We do not assume that the vector perturbations vanish $\omega_i=0$, rather use the fact that the divergence of $\omega_i$ vanishes, $\nabla^i\omega_i=0$. Similarly with the tensor modes $\chi_{ij}$
which are   transverse and traceless. This leads us to use the traceless-longitudinal projection operator
\be
 {\cal P}^{ij}=\delta^{ij}-3\frac{\nabla^i\nabla^j}{\nabla^2}\,
 \ee
to determine  $\Phi$ in terms of $\Psi$ from the Einstein equations.  This procedure should be compared to the lowest order procedure in which one simply uses the $i\ne j$ Einstein equation to determine that $\Phi=\Psi$. We also use only the longitudinal part of the $(i0)$ equations so the contribution of the vector perturbations drops.

As we already mentioned, the energy-momentum tensor is that of a cold pressureless fluid of density $\rho$ and three-velocity vector $\boldsymbol{u}$,
\begin{equation}
T_{\mu \nu}= \rho \,u_\mu u_\nu\, .
\label{em}
\end{equation}
As usually done in the literature and to facilitate the comparisons, we work in the so-called single stream approximation and  set to zero the stress tensor.
The four-velocity vector  $u_\mu=(u_0,\boldsymbol{u})$ is normalized since in our units $c=1$.
Finally, we choose units for which $8\pi G=\kappa^2=1$. With these assumptions, the relevant Einstein equations are (see, for instance, Ref. \cite{BMR})
\bea
(00):&& 3\H^2+ 2\nabla^2 \Psi = \rho\, u_0^2\, ,
\label{g00} \\
(i0):&& \nabla_i \Psi'+  \H \nabla_i\Psi = \frac{1}{2} \rho\, u_0\, u_i\, ,
\label{gi0} \\
 (ij):&& \left(2 \Psi''+6\H \Psi' -2 \left(\H^2-2 \frac{a''}{a}\right)\Psi
+ (\nabla\Psi)^2-\nabla^2\left(\Psi-\Phi\right)\right)\delta_{ij} \cr
 && +\nabla_i\nabla_j\left(\Psi-\Phi\right)-2\nabla_i\Psi \nabla_j\Psi
= \rho \,u_i u_j\, .
\label{gij}
\eea
For future reference we define the (matter) density contrast $\delta\equiv(\rho-\overline{\rho})/\overline{\rho}$, where $\overline{\rho}= 3\H^2$ is the background density. The density contrast can be expressed in terms of the gravitational potential using the $(00)$ equation (\ref{g00}),
\be
\delta=\frac{2\nabla^2\Psi}{3\H^2}.
\label{defdel}
\ee
The traceless-longitudinal part of the $(ij)$ equation is given by
\begin{equation}
{\cal P}^{ij} \left(\nabla_i \nabla_j (\Psi-\Phi) -2 \nabla_i \Psi \nabla_j\Psi\right) ={\cal P}^{ij} \left(\rho\, u_i u_j\right)\, .
\label{inejeq}
\end{equation}
We may evaluate explicitly some of the terms (and divide the whole equation by $- 2$),
\begin{equation}
\nabla^2 (\Psi-\Phi) + \left( \nabla\Psi\right)^2  -3\frac{\nabla^i\nabla^j}{\nabla^2} \left( \nabla_i \Psi\nabla_j\Psi\right) =-\frac{1}{2}\rho \boldsymbol{u}^2+\frac{3}{2}\frac{\nabla^i\nabla^j}{\nabla^2} \left( \rho u_i u_j\right)\, .
\label{inejeq2}
\end{equation}
We will use Eqs.~(\ref{inejeq2}) and (\ref{gi0}) to express $\nabla^2(\Psi-\Phi)$ in terms of $\Psi$ only which will allow us to derive an equation for $\Psi$ that does not involve $\Phi$. Despite appearances, Eq.~(\ref{inejeq2}) is completely local. In fact, all terms have at least two space derivatives acting on the fields.

Let us now look at $1/6$ of the trace of Eq. (\ref{gij})
\begin{eqnarray}
&& \Psi''+3\H \Psi' -  \left(\H^2-2 \frac{a''}{a}\right)\Psi
+ \frac{1}{6}(\nabla\Psi)^2- \frac{1}{3} \nabla^2\left(\Psi-\Phi\right)
= \frac{1}{6} \rho \boldsymbol{u}^2\, .
\label{treq}
\end{eqnarray}
The next step is to substitute Eq.~(\ref{inejeq2}) into Eq.~(\ref{treq}) and obtain
\begin{eqnarray}
 \Psi''+3\H \Psi' -  \left(\H^2-2 \frac{a''}{a}\right)\Psi
+ \frac{1}{2}(\nabla\Psi)^2 - \frac{\nabla^i\nabla^j}{\nabla^2} \left(\nabla_i \Psi\nabla_j\Psi\right)   =\frac{1}{2}\frac{\nabla^i\nabla^j}{\nabla^2} \left(\rho\, u_i u_j\right)\, .
\label{treqfinal}
\end{eqnarray}
Finally, we can  use the $(0i)$ equation (\ref{gi0}) to integrate out the three-velocity $u_i$ from Eq.~(\ref{treqfinal})
\begin{eqnarray}
\Psi''+3\H \Psi' -  \left(\H^2-2 \frac{a''}{a}\right)\Psi
+ \frac{1}{2}(\nabla\Psi)^2 - \frac{\nabla^i\nabla^j}{\nabla^2} \left(\nabla_i \Psi\nabla_j\Psi\right)   \nonumber \\ =2\frac{\nabla^i\nabla^j}{\nabla^2} \left(\frac{ \left(\nabla_i \Psi'+  \H \nabla_i\Psi\right)\left(\nabla_j \Psi'+  \H \nabla_j\Psi\right)}{\rho u_0^2}\right)\, ,
\label{nonpert0}
\end{eqnarray}
and the $(00)$ equation (\ref{g00}) to integrate out the energy density $\rho$
\be
\Psi''+3\H \Psi' -  \left(\H^2-2 \frac{a''}{a}\right)\Psi
+ \frac{1}{2}(\nabla\Psi)^2 - \frac{\nabla^i\nabla^j}{\nabla^2} \left( \nabla_i \Psi\nabla_j\Psi\right)
 =\frac{2}{3\H^2}\frac{\nabla^i\nabla^j}{\nabla^2} \left( \frac{ \left(\nabla_i \Psi'+  \H \nabla_i\Psi\right)\left(\nabla_j \Psi'+  \H \nabla_j\Psi\right)}{1+2\nabla^2\Psi/3\H^2}\right)\, .
\label{nonpert1}
\ee
\noindent
This is our proposal for the full non-perturbative equation for the gravitational potential. We expect Eq.~(\ref{nonpert1}) to provide an accurate description of the non-linear evolution of cosmological perturbations on all relevant scales. Deviations form Eq.~(\ref{nonpert1}) will occur for situations in which our assumptions about the matter content of the universe become inaccurate. In particular, we have assumed that matter is cold and so its intrinsic pressure is small, and that intrinsic dissipation is small.

We will show shortly that Eq.~(\ref{nonpert1})  reproduces the standard perturbative results: that the gravitational potential is constant at the linear order and that it grows with time at second- and third-order
(linearly  and quadratic in  the scale factor, respectively) .  However, even without knowing the exact form of its full solution, Eq.~(\ref{nonpert1}) tells us that $\Psi$ cannot grow indefinitely. Indeed,  when the density contrast grows  with time, that is when $(2\nabla^2\Psi/3\H^2)$ becomes increasingly large, the right-hand side of Eq.~(\ref{nonpert1}) gets more and more   suppressed so  $\Psi$ stops growing and most probably
reaches an attractor solution corresponding to a density contrast that does not evolve in time any longer.

\section{Reconstructing perturbation theory}
\noindent
We now verify that the perturbative solution of Eq.~(\ref{nonpert1}) reproduces the known perturbative solutions for $\Psi$, the density contrast $\delta$ and the velocity $\boldsymbol{u}$.
First,  we linearize Eq.~(\ref{nonpert1}) and obtain the standard equation
\begin{equation}
\Psi''+3\H \Psi' -2  \left(\H^2-2 \frac{a''}{a}\right)\Psi=0\, ,
\end{equation}
whose solution in  matter domination does not evolve with time
\begin{equation}
\Psi(\boldsymbol{x},\tau)=\Psi_{\rm L}(\boldsymbol{x})\, .
\label{psilin}
\end{equation}
For future use we recall that the resulting solution for $\boldsymbol{u}$ is
\begin{equation}
\boldsymbol{u}_{\rm L}=\frac{2\boldsymbol{\nabla} \Psi_{\rm L}}{3\H}= \frac{\tau}{3}\boldsymbol{\nabla} \Psi_{\rm L}\, ,
\label{ulin}
\end{equation}
where in the second equality we have used the explicit form of $\H$ for matter domination, $\H=2/\tau$.
The corresponding solution for $\delta$ can be obtained from its relation to $\Psi$, Eq.~(\ref{defdel}), so
\begin{equation}
\delta_{\rm L}=\frac{1}{6}\tau^2
 \nabla^2 \Psi_{\rm L} =\frac{1}{6} a(\tau) \nabla^2 \Psi_{\rm L}\, .
\label{deltadef}
\end{equation}
Next we solve for $\Psi$ to second-order in perturbation theory. The idea is to substitute in the second-order terms the linear solution. Therefore, we take the non-perturbative Eq.~(\ref{nonpert1}), expand it to second-order in $\Psi$ and take only terms that do not vanish on the constant leading order solution (\ref{psilin}),
\be
 \Psi''+3\H \Psi' - \Psi \left(\H^2-2 \frac{a''}{a}\right)
+ \frac{1}{2}(\nabla\Psi)^2 - \frac{\nabla^i\nabla^j}{\nabla^2} \biggl( \nabla_i \Psi\nabla_j\Psi\biggr)=\frac{2}{3}\frac{\nabla^i\nabla^j}{\nabla^2} \biggl( \nabla_i\Psi   \nabla_j\Psi\biggr)\, .
\label{nonpertII1}
\ee
For matter domination it reduces to
\be
\Psi''+3\H \Psi' =\frac{5}{3}\frac{\nabla^i\nabla^j}{\nabla^2} \biggl(\nabla_i\Psi_{\rm L}  \nabla_j\Psi_{\rm L}\biggr)- \frac{1}{2}(\nabla\Psi_{\rm L})^2 \, ,
\label{nonpertII2}
\ee
whose solution is
\be
\Psi_2= \frac{1}{14} \tau^2 \left[\frac{5}{3}\frac{\nabla^i\nabla^j}{\nabla^2} \biggl(\nabla_i\Psi_{\rm L}   \nabla_j\Psi_{\rm L}\biggr)- \frac{1}{2}(\nabla\Psi_{\rm L})^2\right]\, .
\label{psi2ndorder}
\ee
To evaluate $\Psi_2$ in an explicit way it is easier to calculate $\nabla^2\Psi_2$,
\be
\nabla^2\Psi_2= \frac{1}{14} \tau^2 \left(\frac{5}{3}\nabla^i\nabla^j \left( \nabla_i\Psi_{\rm L}   \nabla_j\Psi_{\rm L}\right)- \frac{1}{2}\nabla^2\left(\nabla\Psi_{\rm L})^2\right)\right)\, .
\label{2ndorder}
\ee
Thus, we  need
\be
\nabla^i\nabla^j \left( \nabla_i\Psi_{\rm L}   \nabla_j\Psi_{\rm L}\right)=
(\nabla^2\Psi_{\rm L})^2+2 \nabla^i\Psi_{\rm L}\nabla^2\nabla_i\Psi_{\rm L}+ \nabla^i\nabla^j\Psi_{\rm L}\nabla_i\nabla_j\Psi_{\rm L}
\ee
and
\begin{eqnarray}
\nabla^i\nabla_i \left( \nabla^j\Psi_{\rm L}   \nabla_j\Psi_{\rm L}\right)=
2 \nabla^i\Psi_{\rm L}\nabla^2\nabla_i\Psi_{\rm L}+2 \nabla^i\nabla^j\Psi_{\rm L}\nabla_i\nabla_j\Psi_{\rm L}\, .
\end{eqnarray}
Substituting into Eq.~(\ref{2ndorder}) we find
\bea
\nabla^2\Psi_2=  \frac{1}{14} \tau^2 \Biggl(\frac{5}{3}(\nabla^2\Psi_{\rm L})^2+ \frac{7}{3} \nabla^i\Psi_{\rm L}\nabla^2\nabla_i\Psi_{\rm L} +\frac{2}{3} \nabla^i\nabla^j\Psi_{\rm L}\nabla_i\nabla_j\Psi_{\rm L}\Biggr)
\label{2ndorder2}
\end{eqnarray}
In momentum space it gives
\begin{eqnarray}
\label{delta2}
 \delta_2(\boldsymbol{k}_3,\tau) &=& -\frac{1}{6}k_3^2\tau^2\Psi_2(\boldsymbol{k}_3,\tau)  \cr  &=& \int\frac{{\rm d}^3 k_1}{(2\pi)^3}\int \frac{{\rm d}^3 k_2}{(2\pi)^3}
\left[\frac{5}{7}  +\frac{1}{2} (\boldsymbol{k}_1\cdot\boldsymbol{k}_2)\frac{k_1^2+k_2^2}{k_1^2 k_2^2}+ \frac{2}{7}\frac{(\boldsymbol{k}_1\cdot\boldsymbol{k}_2)^2}{k_1^2 k_2^2}\right] \\ &\times& \delta_{D}\left(\boldsymbol{k}_1+\boldsymbol{k}_2-\boldsymbol{k}_3\right)\, \delta_{\rm L}(\boldsymbol{k}_1,\tau)
\delta_{\rm L}(\boldsymbol{k}_2,\tau)\, , \nonumber
\end{eqnarray}
which reproduces the standard second-order kernel for the density contrast \cite{pr}. Using Eqs. (\ref{gi0}) and
(\ref{2ndorder2}), it is also straightforward to recover the kernel for the (divergence of the) velocity
$\theta=\boldsymbol{\nabla}\cdot\boldsymbol{u}$.
One expands Eq.~(\ref{gi0}) to second-order, using the time dependence of $\Psi_2$
\be
4\H{\bf \nabla} \Psi_{2}=\overline{\rho}\left(\boldsymbol{u}_{2}+
\delta_{\rm L} \boldsymbol{ u}_{\rm L}\right).
\label{uexpansion}
\ee
Using the lowest order relations, the equation for the divergence of the velocity at second-order becomes
\be
\frac{{\bf \nabla}\cdot\boldsymbol{ u}_{2}}{\H}=2\,\delta_{2}-(\delta_{\rm L})^2
-\boldsymbol{\nabla}\delta_{\rm L} \cdot \frac{{{\boldsymbol{\nabla}} \delta_{\rm L}}}{
\boldsymbol{\nabla}^2}.
\label{uexpansion1}
\ee
In momentum space, using the explicit solution (\ref{2ndorder2}) the final result is,
\bea
-\frac{\theta_2(\boldsymbol{k}_3,\tau)}{\cal H}&=& \int\frac{{\rm d}^3 k_1}{(2\pi)^3}\int \frac{{\rm d}^3 k_2}{(2\pi)^3}
\left[\frac{3}{7}  +\frac{1}{2} (\boldsymbol{k}_1\cdot\boldsymbol{k}_2)\frac{k_1^2+k_2^2}{k_1^2 k_2^2}+ \frac{4}{7}\frac{(\boldsymbol{k}_1\cdot\boldsymbol{k}_2)^2}{k_1^2 k_2^2}\right] \\ &\times & \delta_{D}\left(\boldsymbol{k}_1+\boldsymbol{k}_2-\boldsymbol{k}_3\right) \delta_{\rm L}(\boldsymbol{k}_1,\tau)\,
\delta_{\rm L}(\boldsymbol{k}_2,\tau)\,, \nonumber
\eea
which agrees with the standard Newtonian kernel.
Finally, we compare the third-order solution of our equation to the known solution of SPT.
The third-order equation is given by
\be
\Psi_3''+3\H\Psi_3'=- \left({\bf \nabla}\Psi_2\cdot{\bf \nabla}\Psi_{\rm L}\right)+\frac{14}{3}\frac{\nabla^i\nabla^j}{\nabla^2} \biggl(\nabla_i\Psi_2  \nabla_j\Psi_{\rm L}\biggr)- \frac{4}{9\H^2}\frac{\nabla^i\nabla^j}{\nabla^2} \biggl(\nabla_i\Psi_{\rm L}\nabla_j\Psi_{\rm L} \nabla^2 \Psi_{\rm L}\biggr) \, .
\label{real3rdorder}
\ee
Exposing the explicit time dependence of the second-order solution $\Psi_2=\tau^2\widetilde{\Psi}_2$ and solving for the $\tau$ dependence, we find
\be
\Psi_3=\frac{\tau^4}{36}\left[- \left({\bf \nabla}\widetilde{\Psi}_2\cdot{\bf \nabla}\Psi_{\rm L}\right)+\frac{14}{3}\frac{\nabla^i\nabla^j}{\nabla^2} \biggl(\nabla_i\widetilde{\Psi}_2  \nabla_j\Psi_{\rm L}\biggr)- \frac{1}{9}\frac{\nabla^i\nabla^j}{\nabla^2} \biggl(\nabla_i\Psi_{\rm L}\nabla_j\Psi_{\rm L} \nabla^2 \Psi_{\rm L}\biggr)\right] \, .
\label{real3rdorderI}
\ee
In momentum space,
\be
\Psi_3(\k_4)=\frac{\tau^4}{36}\int\frac{{\rm d}^3 k_1}{(2\pi)^3}\int \frac{{\rm d}^3 k_2}{(2\pi)^3}\int\frac{{\rm d}^3 k_3}{(2\pi)^3}\ f_3(\k_1,\k_2,\k_3)\ \delta_D(\k_4-\k_1-\k_2-\k_3)\ \Psi_{\rm L}(\k_1)\Psi_{\rm L}(\k_2)\Psi_{\rm L}(\k_3) \, .
\label{mom3rdorderI}
\ee
The function $f_3$ is a sum of three terms
\bea
f_3^{(1)}&=& - \frac{1}{14} \k_1\cdot(\k_2+\k_3)\left[\frac{5}{3}\frac{(\k_2+\k_3)\cdot \k_2(\k_2+\k_3)\cdot \k_3}{|\k_2+\k_3|^2}-\frac{1}{2}(\k_2\cdot\k_3)
\right]\, ,\nonumber \\
f_3^{(2)}&=& \frac{1}{3} \frac{(\k_4\cdot\k_1)( \k_4\cdot(\k_2+\k_3))}{k_4^2}
\left[\frac{5}{3}\frac{(\k_2+\k_3)\cdot \k_2(\k_2+\k_3)\cdot \k_3}{|\k_2+\k_3|^2}-\frac{1}{2}(\k_2\cdot\k_3)
\right]\
\, ,\nonumber \\
f_3^{(3)}&=& -\frac{1}{9} \frac{(\k_4\cdot\k_1)( \k_4\cdot\k_2)\ k_3^2}{k_4^2}\, .
\eea
Using the relationship (\ref{deltadef}) between $\delta_{\rm L}$ and $\Psi_{\rm L}$, we obtain
\bea
\label{delta3}
\delta_3(\k_4)&=&-\frac{1}{6}k_4^2\tau^2\Psi_3(\boldsymbol{k}_4,\tau) \nonumber\\
&=&\int\frac{{\rm d}^3 k_1}{(2\pi)^3}\int \frac{{\rm d}^3 k_2}{(2\pi)^3}\int\frac{{\rm d}^3 k_3}{(2\pi)^3}\, F_3 (\k_1,\k_2,\k_3) \delta_D(\k_4-\k_1-\k_2-\k_3) \delta_{\rm L}(\k_1)\delta_{\rm L}(\k_2)\delta_{\rm L}(\k_3) \, ,\nonumber\\
F_3(\k_1,\k_2,\k_3)&=&\frac{f_3(\k_1,\k_2,\k_3)}{k_1^2 k_2^2 k_3^2}.
\eea
We have verified that  this coincides with the third-order kernel for the density contrast \cite{pr}.
We expect that the perturbative series of the solution of our equation is equal to the solution in SPT. This can be verified by comparing the recursion relation of the perturbation series resulting from our equation and the standard kernels of SPT \cite{pr}. We postpone the explicit verification of the all orders relationship between the two kernels to a future investigation.

\section{The mean-field approximation}
\noindent
After verifying that  the solution of non-perturbative equation (\ref{nonpert1})  reproduces the perturbative results at first-, second-, and third-order, we turn to improve the solution in a significant way, using the mean-field approximation.  This will also allow us to make contact with the current techniques of solving for the density contrast beyond standard perturbation theory.

We are interested mainly in the dynamics at small length scales for which linear perturbation theory for the density contrast is not valid.
The idea is to use this separation of scales to average over the slowly varying (in space) large scale velocity fields and produce an improved linear equation for the gravitational potential which will take into account the effective dissipation that these large scale velocity fields induce.

The simplest way to apply the mean-field approximation is to linearize the equation by setting higher order terms to their lowest order classical solutions. Implementing this approximation is more complicated if we start directly with Eq.~(\ref{nonpert1}) because the perturbative orders are mixed there (see however
below). The easiest way to correctly identify the mean-field equation is to start from Eq.~(\ref{gij}). One neglects all the second-order terms on its left-hand side  and sets the right-hand side  to be equal to $\overline{\rho}\delta$ times the average of $\boldsymbol{u}^2$,
\begin{equation}
\Psi''+3\H \Psi' -  \left(\H^2-2 \frac{a''}{a}\right)\Psi=\frac{1}{3}\nabla^2 \Psi \langle \boldsymbol{u}^2\rangle\, ,
\end{equation}
where the average is over the initial conditions $\Psi_{\rm L}$ which is assumed here to be a Gaussian random field. All other terms only contribute a small correction to the background equation.
We may now perform the average and restrict to matter domination, setting $\tau_0=1$,
\begin{equation}
\Psi''+3\H \Psi' =\frac{1}{3}\nabla^2 \Psi\,\langle \boldsymbol{u}^2(\tau_0)\rangle \tau^2\, .
\label{meanfield}
\end{equation}
Alternatively, one can start from Eq. (\ref{nonpert1}) and separate the perturbative orders carefully. First, it is necessary to use the form of the second-order solution Eq.~(\ref{psi2ndorder}) to find that ${\bf \nabla }\Psi_2'= \H{\bf \nabla }\Psi_2$. This is then substituted into Eq.~(\ref{uexpansion}) from which one finds that to second-order,
\be
\nabla_i \Psi'+\H\nabla_i\Psi=\H\nabla_i\Psi_{\rm L} \left(1+\frac{2\nabla^2\Psi_{\rm L}}{3\H^2}\right)\, .
\ee
When this expression is substituted back into Eq.~(\ref{nonpert1}) and reexpressed using Eq.~(\ref{ulin}) in terms of the linear order velocity $\boldsymbol{u}_{\rm L}$, one obtains after averaging exactly the mean-field equation  Eq. (\ref{meanfield}).

As we have discussed above, the mean-field approximation relies on scale separation. To understand this let us express the non-linear term in momentum space,
\bea
\left(\nabla^2 \Psi\  \boldsymbol{u}^2\right)_{\boldsymbol{k}}(\tau_0)&=&-\int\frac{{\rm d}^3 q_{1}}{(2\pi)^3}\int\frac{{\rm d}^3 q_2}{(2\pi)^3}\int\frac{{\rm d}^3 q_3}{(2\pi)^3}\,q_1^2\,\Psi_{\boldsymbol{q}_1}\,\boldsymbol{u}_{\boldsymbol{q}_2}\cdot \boldsymbol{u}_{\boldsymbol{q}_3}\,
\delta_D\left(\boldsymbol{k}-\boldsymbol{q}_1-\boldsymbol{q}_2-\boldsymbol{q}_3\right)\nonumber\\
&=&
-\int\frac{{\rm d}^3 q_2}{(2\pi)^3}\int\frac{{\rm d}^3 q_3}{(2\pi)^3}\,\left|\boldsymbol{k}-\boldsymbol{q}_2-\boldsymbol{q}_3\right|^2\,\Psi_{\boldsymbol{k}-\boldsymbol{q}_2-\boldsymbol{q}_3}\,\boldsymbol{u}_{\boldsymbol{q}_2}\cdot \boldsymbol{u}_{\boldsymbol{q}_3}\, .
\label{mfk1}
\eea
If we now restrict the integral to the region for which $k\gg q_2,q_3$ (the ``large-momentum" limit) then
\bea
\left(\nabla^2 \Psi\  \boldsymbol{u}^2\right)_{\boldsymbol{k}}(\tau_0)
&\simeq&
-k^2\Psi_{\boldsymbol{k}} \int\frac{{\rm d}^3 q_2}{(2\pi)^3}\int\frac{{\rm d}^3 q_3}{(2\pi)^3}\,\,\boldsymbol{u}_{\boldsymbol{q}_2}\cdot \boldsymbol{u}_{\boldsymbol{q}_3}\, ,
\label{mfk2}
\eea
which in the mean-field approximation gets replaced by
\be
-k^2\Psi_{\boldsymbol{k}} \Big\langle\int\frac{{\rm d}^3 q_2}{(2\pi)^3}\int\frac{{\rm d}^3 q_3}{(2\pi)^3}\,\,\boldsymbol{u}_{\boldsymbol{q}_2}\cdot \boldsymbol{u}_{\boldsymbol{q}_3}\Big\rangle=
-k^2\Psi_{\boldsymbol{k}}\,\int\frac{{\rm d}^3 q}{(2\pi)^3}\,\boldsymbol{u}_{\boldsymbol{q}}\cdot \boldsymbol{u}_{-\boldsymbol{q}}=
-k^2\Psi_{\boldsymbol{k}}\langle \boldsymbol{u}^2(\tau_0)\rangle\, .
\label{mfk3}
\ee
The operation of averaging is justified by the fact that the high-frequency mode $\Psi_{\k}$ ``sees" the other slower modes as a approximate constant background which in turn  is equivalent to  a spatial averaging (and to an ensemble averaging due to the ergodic theorem).
Notice that the same result can be obtained starting from Eq. (\ref{treqfinal}). In such a case the term in the right-hand side of that equation is the only one retained and gives  in momentum space and in the high-momentum limit (at $\tau=\tau_0$)
\bea
\frac{4}{9{\cal H}^2}\Psi_{\boldsymbol{k}} \Big\langle\int\frac{{\rm d}^3 q_2}{(2\pi)^3}\int\frac{{\rm d}^3 q_3}{(2\pi)^3}\,\,
\left(\boldsymbol{k}\cdot \boldsymbol{q}_2\right)\left(\boldsymbol{k}\cdot \boldsymbol{q}_3\right)\Psi_{\boldsymbol{q}_2}\Psi_{\boldsymbol{q}_3}
\Big\rangle&=&- \frac{4}{9{\cal H}^2}\Psi_{\boldsymbol{k}} \int\frac{{\rm d}^3 q}{(2\pi)^3}\,\,
\left(\boldsymbol{k}\cdot \boldsymbol{q}\right)^2
\Psi_{\boldsymbol{q}}\Psi_{-\boldsymbol{q}}\nonumber\\
&=&-\frac{4}{9{\cal H}^2}\frac{1}{3}k^2\Psi_{\boldsymbol{k}} \int\frac{{\rm d}^3 q}{(2\pi)^3}\,\,
q^2\,
\Psi_{\boldsymbol{q}}\Psi_{-\boldsymbol{q}}\nonumber\\
&=&-\frac{1}{3}k^2\Psi_{\boldsymbol{k}}\langle \boldsymbol{u}^2(\tau_0)\rangle\, .
\eea
The quantity $\sigma_u^2\equiv\frac{1}{3} \langle \boldsymbol{u}^2(\tau_0)\rangle$ is related in a simple way to a quantity that is widely used in the literature,
\be
\sigma_v^2=\frac{1}{4}\sigma_u^2\, ,
\ee
where
\be
\sigma_v^2\equiv\frac{1}{3} \int \frac{{\rm d}^3 q}{q^2} \, P_\delta^{\rm L}(q)\, ,
\ee
with $P_\delta^{\rm L}$ being the linear power spectrum of the density contrast.
To facilitate the comparison of our results to the literature we will use $\sigma_v^2$ in what follows.

Substituting the explicit time-dependence of $\H$ and going to momentum space we find from Eq.~(\ref{meanfield})
\begin{equation}
\Psi_{\boldsymbol{k}}''+\frac{6}{\tau} \Psi_{\boldsymbol{k}}' =-4 k^2 \sigma_v^2 \tau^2 \Psi_{\boldsymbol{k}}\, .
\label{explicitmeanfield}
\end{equation}
To solve Eq.~(\ref{explicitmeanfield}) we need to impose boundary conditions. We choose to impose boundary conditions that assume that the solution is exactly equal to the constant perturbative solution until time $\tau_0$,  so $\Psi_{\boldsymbol{k}}(\tau_0)=\Psi_{\rm L}$ and $\Psi_{\boldsymbol{k}}'(\tau_0)=0$. Then for times $\tau>\tau_0$ the non-linear effects are ``turned-on" and start to affect the solution.
The solution of Eq.~(\ref{explicitmeanfield}) with the specified boundary conditions is the confluent hypergeometric function,
\begin{eqnarray}
\Psi_{\k}(\tau)=\Psi_{\boldsymbol{k}}(\tau_0)\ {}_0F_1\left[\frac{9}{4},-\frac{1}{4} k^2 \sigma_v^2 (\tau^4-\tau_0^4)\right]
\end{eqnarray}
and  can be expressed in terms of a Bessel function
\begin{equation}
\label{bessel}
 \left(\frac{1}{4} k^2 \sigma_v^2 (\tau^4-\tau_0^4)\right)^{-5/8}\ \Gamma\left(\frac{9}{4}\right)\ J_{5/4}\left[
       2 k\sigma_v (\tau^2-\tau_0^2)\right]\, .
\end{equation}
The solution starts at unity for $\tau=\tau_0$, decays exponentially for a while and then oscillates within an envelope that decays as $\tau^{-7/2}$  with a  periodicity that increases as $\tau^2$.
The relevant information about the solution is essentially contained in the solution of the first-order  equation
\begin{equation}
3\H \Psi' =4 \nabla^2 \Psi\ \sigma_v^2\tau^2\, ,
\label{1meanfield}
\end{equation}
which is
\begin{eqnarray}
\left(\Psi_{\rm MF}\right)_{\boldsymbol{k}}(\tau)=\Psi_{\boldsymbol{k}}(\tau_0)\exp\left({-\frac{1}{6} k^2 \sigma_v^2 (\tau^4-\tau_0^4)}\right)
\label{meanfieldsol}
\end{eqnarray}
in the region in which the argument of the exponential is less than about unity.
The reason why the solution~(\ref{meanfieldsol}) suffices is that the validity of the mean-field solution itself is restricted to times $\tau$ and wavenumbers $k$ such that the exponent is less than or order unity. This can be seen by looking at Eq.~(\ref{gi0}). To derive the mean-field approximation we have used the lowest order solution for which $\Psi'=0$. However, to mean-field solution itself is time dependent, so our estimate of the average $\langle \boldsymbol{u}^2\rangle$ will no longer be accurate when the solution deviates substantially from a constant.
Let us compare $\Psi'$ to $\H \Psi$ for the mean-field solution  (\ref{meanfieldsol}) when $\tau\gg\tau_0$,
\begin{eqnarray}
\Psi'&=&-\frac{2}{3} k^2 \sigma_v^2 \tau^3 \Psi\, ,\nonumber\\
\H\Psi&=& \frac{2}{\tau} \Psi\, .
\end{eqnarray}
They  become comparable when $
\frac{1}{3} k^2 \sigma_v^2 \tau^4 \sim 1$.
This means that the exponential solution is completely sufficient in the region of validity of the solution. For instance, at redshift $z=0$, the exponential decay is valid at least up to
scales $k\sim 0.25\,h\,{\rm Mpc}^{-1}$ and then it gets replaced by a power-law decay. A similar  estimate for the region of validity of the solution can be obtained by comparing $\Psi''$ to respect to $3\H \Psi'$ in Eq. (\ref{explicitmeanfield}).

Even though we cannot trust the exponential decay or the solution at times and/or wavenumbers
for which $k^2 \sigma_v^2 \tau^4 \gg  1$, we do expect the softening of the decay to persist for the true solution of the full non-perturbative equation. The reason is that we expect the velocities to decay as a result of the decreased gravitational potential, and when the velocities decay, the effective dissipation that they induce will be weaker, so the rate of decay of the gravitational potential will be smaller.

The mean-field approximation is also not expected to be valid for combination of times and wavenumbers
for which $k^2 \sigma_v^2 \tau^4 \ll 1$. The reason is that the  scale separation that we have invoked previously to perform the averaging operation is not valid. Under such conditions the expected change to the solution is that,  instead of being described by an expansion of the form $e^{-\frac{1}{2} k^2 \sigma_v^2 \tau^4}=1-\frac{1}{2} k^2 \sigma_v^2 \tau^4+\cdots$,  the solution will be modified to $1-c_1 k^2 \sigma_v^2 \tau^4+\cdots$ where $c_1\ne \frac{1}{2}$. The modified coefficient $c_1$ can be determined by a perturbative (one-loop) computation.

It is possible, if one so wishes, to iteratively improve the mean-field approximation by using the solution at each given step to recalculate $\langle \boldsymbol{u}^2\rangle$ for the next step, generating an expansion in $\exp\left(- k^2 \sigma_v^2\tau^4/6\right)$.  Another improvement would be in matching the mean-field solution in a better way to the perturbative solution or combining both improvements. We will not pursue these possible improvements any further, leaving them to future research.

\subsection{The mean-field solution for the density contrast}
\noindent
The density contrast and the gravitational potential are related via Eq.~(\ref{g00}). One would naively think that to obtain the mean-field solution for $\delta$ it is enough to substitute the exponential solution~(\ref{meanfieldsol}) into the relation between $\Psi$ and $\delta$ and obtain the desired result. However, this is not true, as we now show.

From Eq.~(\ref{defdel}) we find that the mean-field equation that $\delta$ obeys is somewhat different the its counterpart Eq. (\ref{meanfield}),
\be
\delta''+\H \delta' -\frac{3}{2}\H^2 \delta=\frac{1}{3} \langle \boldsymbol{u}^2\rangle \nabla^2 \delta \, .
\label{delmf}
\ee
In momentum space, substituting the explicit time dependence of matter domination,
\be
\delta_{\k}''+\frac{2}{\tau} \delta_{\k}'-\frac{6}{\tau^2} \delta_{\k}=-4 k^2 \sigma_v^2 \tau^2 \delta_{\k}\, .
\label{delmfmom}
\ee
The different coefficient of $\H$ rather than $3\H$ in front of the friction term in Eq.~(\ref{delmf}) causes the time dependence of the mean-field solution for $\delta$ to change. The full solution of Eq.~(\ref{delmfmom}) can again be expressed in terms of Bessel functions similar to the solution (\ref{bessel}).
 The full solution is subject to the same restrictions as before so the relevant information is contained in the approximate solution that is obtained by multiplying the linear solution $\delta_{\rm L}$ which solves the equation
\be
\delta_{\rm L}''+\frac{2}{\tau} \delta_{\rm L}'-\frac{6}{\tau^2} \delta_{\rm L}=0\, ,
\ee
 by the solution of the first-order equation
\be
\frac{2}{\tau} \widetilde{\delta}_{\k}^{\ \prime}=-4 k^2 \sigma_v^2 \tau^2 \widetilde{\delta}_{\k}\, ,
\label{delmfmomI}
\ee
whose solution is
\be
\widetilde{\delta}_{\k}=\widetilde{\delta}_{\k}(\tau_0)e^{-\frac{1}{2} k^2 \sigma_v^2 (\tau^4-\tau_0^4)}\, .
\label{delmfmomII}
\ee
So the final result for the mean-field solution is
\be
\left(\delta_{\rm MF}\right)_{\k}=(\delta_{\rm L})_{\k} e^{-\frac{1}{2} k^2 \sigma_v^2 (\tau^4-\tau_0^4)}
\label{delmfmomFin}
\ee
An alternative way to find the mean-field solution for $\delta$ is by starting from the standard fluid equations for $\delta$ and  $\theta\equiv \boldsymbol{\nabla}\cdot\boldsymbol{u}$,
\begin{eqnarray}
\label{delt1}
\delta'+ (1+\delta)\ \theta +u^\alpha\nabla_\alpha\delta&=&0\, ,\\
\theta'+\H\theta +\frac{3}{2}\H^2\delta + \nabla_\beta u^\alpha \nabla_\alpha u^\beta + u^\alpha\nabla_\alpha \theta&=&0\, .
\label{thet1}
\end{eqnarray}
We then take the time derivative of Eq.~(\ref{delt1})
\begin{eqnarray}
&& \delta''+(1+\delta)\theta' +\theta \delta' +u^\alpha\nabla_\alpha\delta'+(u^\alpha)'\nabla_\alpha\delta=0\, ,
\end{eqnarray}
and substitute  Eq.~(\ref{thet1})
\be
 \delta''-(1+\delta)\left(\H\theta +\frac{3}{2}\H^2\delta + \nabla_\beta u^\alpha \nabla_\alpha u^\beta + u^\alpha\nabla_\alpha \theta\right) +\theta \delta'+(u^\alpha)'\nabla_\alpha\delta +u^\alpha\nabla_\alpha\delta'=0\, .
\ee
Now we substitute again Eq.~(\ref{delt1}) in the form $\H\theta=-\H\delta'-\H\delta\theta\ -\H u^\alpha\nabla_\alpha \delta$ and find after some straightforward algebraic manipulations,
\begin{eqnarray}
&& \delta''+ \H\delta' -\frac{3}{2}\H^2\delta \cr && - \nabla_\beta u^\alpha \nabla_\alpha u^\beta - u^\alpha\nabla_\alpha \theta  -\frac{3}{2}\H^2\delta^2 -\delta \nabla_\beta u^\alpha \nabla_\alpha u^\beta\nonumber\\
&& -\delta u^\alpha\nabla_\alpha \theta +\left(\H u^\alpha+(u^\alpha)'\right)\nabla_\alpha\delta -\left(\theta+u^\alpha\nabla_\alpha\right)\left[(1+\delta)\ \theta +u^\alpha\nabla_\alpha\delta\right]=0\, .
\end{eqnarray}
To identify the mean-field solution we now go to the ``high-momentum" limit,  which in this case means taking the highest power of gradients acting on $\delta$;  the equation simplifies enormously. The leading term comes from the second derivative term  $u^\alpha\nabla_\alpha u^\beta\nabla_\beta\delta$, and one takes only the terms where the gradients act on $\delta$,
\be
 \delta''+ \H\delta' -\frac{3}{2}\H^2\delta -u^\alpha u^\beta\nabla_\alpha \nabla_\beta\delta=0\, .
\ee
Solving this equation in the mean-field approximation,
\be
\delta''+ \H\delta' -\frac{3}{2}\H^2\delta = \langle u^\alpha u^\beta\rangle\nabla_\alpha \nabla_\beta\delta =
\frac{1}{3} \langle \boldsymbol{u}^2\rangle\ \nabla^2\delta\, .
\ee
which is the same equation as we have found for $\delta$ previously, Eq.~(\ref{delmf}). The solutions of the two identical
equations are, of course, the same.

\section{Comparison to the RPT and RG methods}
\noindent
The exponential suppression that we have obtained in the mean-field approximation
matches perfectly the exponential suppression obtained for the non-linear propagator (to be discussed shortly)
within the RPT proposed
in Ref. \cite{RPT} and the RG method put forward in Refs. \cite{MP,MP1}. We would like now to show that this is not a  coincidence, but that the mean-field approximation applied to our non-perturbative
Eq.~(\ref{nonpert1}) captures the essential physics in the large-momentum limit described within the RPT and RG methods.

The two methods   introduce  an object that is called the ``non-linear propagator" which replaces the standard perturbative propagator (the Green's function) and encodes the information about the initial conditions, thus significantly improving over the standard perturbation theory. In the diagrammatic approach of RPT
the non-linear propagator is obtained through   the resummation of an infinite class of diagrams, while in the RG method
it is evaluated by solving  renormalization group equations obtained by introducing a running ultraviolet cut-off.

Let us recall the definition of the non-linear propagator,
\be
G_{ab}(\k,\tau)=g_{ab}(\tau)+\sum\limits_{n=2}^\infty \left\langle \frac{ \delta \Phi_a^{(n)}(\k,\tau)}{\delta\phi_b({\k})}\right\rangle ,
\label{nlprop}
\ee
where the indices $a,b=1,2$ correspond to the different type of perturbations $\phi_1=\delta$ and $\phi_2=\theta$, $g_{ab}$ is the linear propagator, $\Phi^{(n)}$ denotes the $n$'th order expansion of the full solution $\Phi$ in powers of the  linear perturbation $\phi_{1,2}$ and the average is an ensemble average over the initial conditions.

In the explicit calculations the non-linear propagator is calculated to third-order (``one-loop") in the small momentum limit, while
in the large-momentum limit it was found to have an exponential form after resumming  an infinite set of diagrams.
In particular, if
\be
\Phi^{(n)}=\int\,{\cal F}_{ab a_1\cdots a_n}^{(n)}(\boldsymbol{k}_1,\cdots,\boldsymbol{k}_n,\tau)\,\delta_D(\boldsymbol{k}-\boldsymbol{k}_1-\cdots -\boldsymbol{k}_n)\,
\phi_{a_1}(\boldsymbol{k}_1)\cdots\phi_{a_n}(\boldsymbol{k}_n)\,
\ee
is the series expansion in terms of the linear $\phi$'s and of the standard perturbation theory kernels
${\cal F}_{a a_1\cdots a_n}^{(n)}$, the non-linear propagator may be written as the infinite sum
(corresponding to an infinite number of diagrams)
\cite{RPT}
\be
G_{ab}(k,\,\tau)=g_{ab}(\tau)+\sum_{n=1}^\infty\,(2n+1)!!\int\, {\cal F}_{ab}^{(2n+1)}(\boldsymbol{k},\boldsymbol{k}_1,-\boldsymbol{k}_1,\cdots,\boldsymbol{k}_n,-\boldsymbol{k}_n,\tau)P_{\rm L}^\delta(k_1)\cdots P_{\rm L}^\delta(k_n)\, ,
\ee
with
\be
 {\cal F}_{ab}^{(2n+1)}(\boldsymbol{k},\boldsymbol{k}_1,-\boldsymbol{k}_1,\cdots,\boldsymbol{k}_n,-\boldsymbol{k}_n,\tau)= {\cal F}_{aba_1\cdots a_{2n}}^{(2n+1)}(\boldsymbol{k},\boldsymbol{k}_1,-\boldsymbol{k}_1,\cdots,\boldsymbol{k}_n,-\boldsymbol{k}_n,\tau) u_{a_1}\cdots u_{a_{2n}}\, ,\,\,
 u_{a_i}=(1,1)\, .
 \ee
 These expressions considerably simplify in the large-momentum limit for which
 \be
 \left\langle \frac{ \delta \Phi_a^{(2n+1)}(\k,\tau)}{\delta\phi_b({\k})}\right\rangle
=g_{ab}(\tau)\frac{(a(\tau)-1)^{2n}}{(2n)!}\int\, {\rm d}^3 q_{2n}\cdots
{\rm d}^3 q_1\,\frac{k^{2n}}{q_{2n}\cdots q_1}\,x_{2n}\cdots x_1\,\langle\delta_{\rm L}(\boldsymbol{q}_{2n})\cdots
\delta_{\rm L}(\boldsymbol{q}_1)\rangle\,, x_i=\frac{\boldsymbol{k}\cdot\boldsymbol{q}_i}{k q_i}\, ,
\ee
such that
\bea
\label{rptg}
G_{ab}(k,\tau)&=&g_{ab}(\tau)+g_{ab}(\tau)\sum_{n=1}^\infty\,\frac{(2n-1)!}{(2n)!}\left[-k^2\sigma_v^2
(a(\tau)-1)^2\right]^{2n}\nonumber\\
&=&
g_{ab}(\tau)\,{\rm exp}\left(-k^2\sigma_v^2
(a-1)^2/2\right)\, , \,\,{\rm large}\,\,k{\rm -limit}.
\eea
This exponential  form of the non-linear propagator in the large momentum limit  can be matched to the full one-loop expression
and then used to calculate the matter power spectrum (and other cosmological observables) for the whole range of momentum (large and small),
 \be
 \label{final}
 P_\delta(k,\tau)=G^2_\delta(k,\tau)P_\delta^{\rm L}(k)+P_{\rm MC}(k,\tau)\, ,
 \ee
where $G_\delta=G_{11}+G_{12}$ and
all contributions which are proportional to the initial
spectrum of fluctuations are now included in the first term, which accounts for the remainder of the primordial power at a given scale after non-linear evolution, and thus has direct information on the linear power spectrum; by contrast, the second term $P_{\rm MC}$ represents the power generated by mode-coupling at smaller scales, and depends on the linear power at different scales than $k$ through complicated convolutions. The important aspect of $P_{\rm MC}$ is  that its time dependence is also dictated by the non-linear propagator. As a result, the convergence  of the perturbation series for $P_{\rm MC}$ is drastically improved. The RG technique of Ref. \cite{MP} offers an alternative, and to some extent, more compact way of getting the same result.

Comparing the definition of the non-linear propagator to our discussion of the mean-field approximation in momentum-space, Eqs~.(\ref{mfk1}--\ref{mfk3}) for the gravitational potential and Eq. (\ref{delmfmomFin}) for the density contrast,  it is clear the in the large-momentum limit, the non-linear propagators are exactly equal to the mean-field solution of our equation.
For instance, in our case the non-linear propagator for $\Psi$ is
\be
G_{\Psi}(\k,\tau)=1+\sum\limits_{n=2}^\infty \left\langle \frac{ \delta \Psi_{n}(\k,\tau)}{\delta\Psi_{\rm L}({\k})}\right\rangle ,
\label{nlproppsi}
\ee
Consider now the non-linear term in the equation in the leading order approximation when all fields are replaced by the lowest order fields $\Psi_{\rm L}$,
\bea
&&-\int\frac{{\rm d}^3 q_{1}}{(2\pi)^3}\int\frac{{\rm d}^3 q_2}{(2\pi)^3}\int\frac{{\rm d}^3 q_3}{(2\pi)^3}\,q_1^2\,\Psi_{\boldsymbol{q}_1}\,u_{\boldsymbol{q}_2}\,u_{\boldsymbol{q}_3}\,\delta_D\left(\boldsymbol{k}-\boldsymbol{q}_1-\boldsymbol{q}_2-\boldsymbol{q}_3\right)\nonumber\\
&=&
\frac{\tau^2}{9}\int\frac{{\rm d}^3 q_{1}}{(2\pi)^3}\int\frac{{\rm d}^3 q_2}{(2\pi)^3}\int\frac{{\rm d}^3 q_3}{(2\pi)^3}\,q_1^2\ \boldsymbol{q}_2 \boldsymbol{q}_3 \,\Psi_{\rm L}(\boldsymbol{q}_1)\,\Psi_{\rm L}(\boldsymbol{q}_2)\,\Psi_{\rm L}(\boldsymbol{q}_3)\,\delta_D\left(\boldsymbol{k}-\boldsymbol{q}_1-\boldsymbol{q}_2-\boldsymbol{q}_3\right),
\label{nlpsi1}
\eea
where we have substituted the expression for the first-order velocity $\boldsymbol{ u}_{\rm L}$ from Eq.~(\ref{ulin}).
In the large-momentum limit,  taking  the functional derivative with respect to $\Psi_{\rm L}(\boldsymbol{q}_1)$ and
averaging  over the remaining two variables,  leads to the mean-field result for the gravitational potential. Of course, a similar
procedure shows that the mean-field solution is equal to    the non-linear propagator for $\delta$ and $\theta$ in the large-momentum
limit.

When we solve Eq.~(\ref{nonpert1}) in the mean-field approximation we first (space) average some of the terms of the equation and then find the solution of the resulting linear equation. In the relevant region of parameter space the equation is first order in the time derivative and so the solution is an exponential of the averaged term. In the RPT method the order is different, as is made particulary clear when using the so-called $\alpha$ method
\cite{amethod}. There one solves a linear equation which is first order in the time derivative for the propagator (or other correlation functions). The solution of this equation is an exponential, which is then averaged over the initial conditions. The two procedures do not necessarily commute and so we do expect differences at some level.

To calculate perturbative corrections with the non-linear propagator one needs to know its full momentum dependence and
not just its large-momentum form. The proposal of the RPT and RG methods is to use the one-loop perturbative
propagator in the low $k$ limit, the exponential form in the large $k$ limit and in intermediate scales
use a specific interpolating function between the two limits. The use of the interpolating function is well motivated physically and matches the data and simulations well.  Clearly, if we use the same approximations as proposed in RPT and  RG,
then the results that we obtain for the cosmological observables will be identical.
The final form (\ref{final}) for the matter power-spectrum in our approach will correspond to
the sum of the homogeneous non-perturbative equation in the mean-field approximation
(corresponding to the first term in Eq.~(\ref{final}))
and the particular  solution induced by  the other non-linear terms of the equation (corresponding to
$P_{\rm MC}$) and containing the mean-field exponentially-suppressed propagator.
This has the effect of taming the bad ultraviolet  behaviour of the solution for $\Psi$ at, for instance, one-loop.  This step exactly follows the proposals of the RPT and RG methods.

The interpolation procedure that connects the high-momentum limit to the perturbative, low-momentum limit, is not based on first principles as are the two limits. One advantage of our non-perturbative Eq. (\ref{nonpert1}) is that it allows to estimate the deviations of the non-linear propagator from the exact Green's function. For example, we have seen that the exponential form of the solution is restricted to times and scales such that the argument of the exponent is not much larger than unity. This is related to the fact that the true solution is a solution of a second-order equation while the exponential solution solves a first-order equation. In the RPT or RG methods, the exponential form of non-linear propagator is derived in the high-momentum approximation. In this limit the equation that the non-linear propagator obeys becomes a first order equation. It would be interesting to understand in detail the connection between our approach and the RPT and RG approaches in this context.

\section{Qualitative analysis of non-perturbative solutions}

The most significant qualitative feature that we can attribute to the exact solutions of our equation is that under generic conditions they do not allow the density contrast to grow without limit. This leads us to expect that in the full solutions the density contrast will saturate asymptotically to an attractor solution which is independent on time and so can only be space-dependent. Since we are always using comoving coordinates, this expectation does not contradict the theoretically well established (and observed) fact that local densities as observed by static observers can grow and become very large.

To understand this point in a more quantitative way, let us look for
a self-similar solution \cite{peebles} of Eq.~(\ref{nonpert1}). The advantage of the self-similar ansatz is that it reduces the partial differential equation~(\ref{nonpert1}) to an ordinary (non-linear) differential equation which is much easier to solve. Here we restrict the discussion to a matter dominated universe. Since there is no preferred scale in the dynamics of a self-gravitating pressureless perfect fluid, cosmological fields should scale with a self-similarity variable given some appropriate initial conditions.
If we express  the gravitational potential $\Psi(\boldsymbol{x},\tau)$ as
\be
\Psi(\boldsymbol{x},\tau)=\tau^\gamma\,\psi(\boldsymbol{y})\, ,\,\,\,\boldsymbol{y}=\boldsymbol{x}/\tau^\nu\, ,
\ee
then $\Psi''\sim \tau^{\gamma-2}$, $\H\Psi'\sim \tau^{\gamma-2}$ and ${\bf \nabla}\Psi\sim \tau^{-\nu}$. To make each of the terms in Eq.~(\ref{nonpert1}) scale with the same power of time, the power-law coefficients $\gamma$ and $\nu$ must obey the relation
\be
\gamma=2(\nu-1)\, .
\ee
In this case the density contrast $\delta=2\nabla^2\Psi/3\H^2$ (See Eq.~(\ref{defdel})) scales as a constant in time.  The scaling of the velocity can be read off of Eq.~(\ref{gi0}),
$\boldsymbol{u}\sim \tau^{\nu-1}$ in complete agreement with the known scaling solution.

To fix the form of the scaling solution completely, that is to determine the parameter $\nu$, additional input is needed. For example, it can be fixed by matching the matter power spectrum resulting from the scaling solution to the linear one. However, this procedure is not as straightforward as it seems because the initial conditions typically do not obey the scaling relations that the scaling solution obeys.
The gravitational potential is a random variable on which one may impose initial conditions only statistically, {\it e.g.} by specifying the linear power spectrum for Gaussian initial conditions. A particular realization of the ensemble of initial conditions will not obey spatial scaling \cite{nonself}. Furthermore, the introduction of (time-independent) smoothing scales and/or cut-off scales breaks self-similarity. Nevertheless, studying self-similar solutions
is useful to confirm our expectation that on non-linear scales the density contrast saturates to a limiting value.

An additional solution to our equation can be found using a simple ansatz
\begin{equation}
\Psi(\boldsymbol{x},\tau)= \Psi_0(\boldsymbol{x}_0 )+A(\boldsymbol{x}_0)\frac{ (\boldsymbol{x}-\boldsymbol{x}_0)^2}{a(\tau)}\, .
\label{ansatz1}
\end{equation}
The velocity $\boldsymbol{u}$ vanishes for this solution; indeed, the $(i0)$ Einstein equation can be expressed as
\begin{equation}
\frac{2}{a(\tau)}\left(a(\tau) \boldsymbol{\nabla}\Psi\right)'=\rho \boldsymbol{u}\, ,
\label{solsc}
\end{equation}
so the chosen ansatz corresponds to a vanishing velocity.
Since Eq.~(\ref{nonpert1}) is non-linear the parameter $A$ in the ansatz (\ref{ansatz1}) is determined by the equation itself. 
It is possible to check that if $A=-1$ the solution satisfies exactly Eq.~(\ref{nonpert1}). When we calculate the value of $\delta$ using Eq.~(\ref{defdel}) we find that it corresponds to a constant density contrast $\delta=-1$, which is nothing but a void. This, in some sense, is a trivial solution whose existence should have been anticipated. However, the way it comes out of the equation is quite non-trivial and reassures us about the correctness of Eq.~(\ref{nonpert1}) and strengthens our expectations that the density contrast cannot grow in time without bounds.

Assuming various other forms for the solution, one can find additional candidate attractor solutions. For example, another candidate solution can be found by using a wave-like ansatz
\be
\Psi=A \ln\left( \boldsymbol{B}\cdot \boldsymbol{x}-\frac{\tau}{\tau_0}\right).
\label{logsol}
\ee
Here $\boldsymbol{B}$ and $\tau_0$ are constants.
To understand the ansatz (\ref{logsol}) better, let us look at its one dimensional version (taking the argument of the logarithm as dimensionless)
\be
\Psi=A \ln(x-v\tau)\, .
\label{logsol1}
\ee
Using Eq.~(\ref{defdel}) we find
\be
\delta=-\frac{A}{6} \tau^2  \frac{1}{(x-v\tau)^2}\, .
\label{dellog}
\ee
When $(x-v\tau)$ becomes small the magnitude of the density contrast grows and in this case we can approximate $\delta\sim \rho$.
Using the same approximation we find 
\be
u=-\frac{12 v}{\tau^2}\, ,
\label{ulog}
\ee
so,  for large $\tau$,  $u$ decreases towards zero.
Combining Eqs.~(\ref{dellog}) and (\ref{ulog}) we find that the right-hand side  of Eq.~(\ref{nonpert1}) is subleading compared to others for large times.

Further, $\Psi''$ scales as $1/(x-v\tau)^2$ while $\Psi'\sim 1/(x-v\tau)$ and is therefore subleading.  Collecting the relevant terms, Eq.~(\ref{nonpert1})  reduces to
\be
\Psi''-\frac{1}{2}(\partial_x \Psi)^2=0\, ,
\label{approxnonpert}
\ee
which fixes the value of $A$ in Eq.~(\ref{logsol1}),
\be
A=-2 v^2\, .
\label{logA}
\ee
This gives the final result for the density contrast (and so for the density)
\be
\rho\sim\frac{1}{3}  \frac{v^2\tau^2}{(x-v\tau)^2}\, .
\label{dellogfinal}
\ee
The solution~(\ref{dellogfinal}) is reminiscent of the form of the density contrast in the Zel'dovich approximation \cite{zeldovichRMP},

\be
\rho(\boldsymbol{x},\tau)\sim \frac{1}{1-a(\tau)\lambda_1(\boldsymbol{q})}\ \frac{1}{ 1-a(\tau)\lambda_2(\boldsymbol{q})}\  \frac{1}{1-a(\tau)\lambda_3(\boldsymbol{q}) }\, ,
\ee
where $\boldsymbol{x}$ is the Eulerian coordinate and $\boldsymbol{q}$ is the Lagrangian one. The sign of the eigenvalues $\lambda_i$ determines the nature of the solution. Comparing the Zel'dovich approximation to the solution~(\ref{dellogfinal}),  we see that this solution corresponds to two equal positive eigenvalues and a vanishing third eigenvalue, corresponding to a collapse to a one dimensional filament.

The full significance of the solutions that we have just discussed and of other non-perturbative solutions to our equation needs, obviously, to be understood in much more detail. We expect to carry this line of investigations to completion in the future.

\section{Conclusions and perspectives}
\noindent
In this paper we have derived a non-perturbative equation for the gravitational potential in the case that the cosmic fluid consists of non-relativistic pressureless particles. Our equation reproduces the known perturbative results (verified explicitly up to third order in perturbation theory) and provides an alternative derivation
of the exponential suppression of the non-linear propagator in the high-momentum limit obtained recently with the RPT and RG methods.

We expect that our current findings can be extended in various directions. First,  a quantitative comparison with observations. In this paper we were not pursuing a full quantitative
computation of cosmological observables. This quantitative analysis and its comparison with data is of utmost importance as it is  the main motivation for the study of non-linear cosmological perturbations. We also expect to be able to extend our results to the cases of cosmological constant (or dark energy)  dominated universe and non-Gaussian initial conditions.

We expect to be able to iteratively improve the mean-field approximation by using the solution at each given step to recalculate $\langle \boldsymbol{u}^2\rangle$ for the next step, generating an expansion in $\exp\left(- k^2 \sigma_v^2\tau^4/6\right)$. This will allow us to extend the region of validity of the solution to larger values of $k^2 \sigma_v^2\tau^4$.  Another improvement would be achieved by improving the matching of the mean-field solution to the perturbative solution which is valid for small values of $k^2 \sigma_v^2\tau^4$ or combining both type of  improvements.

Finding exact non-perturbative solutions or approximate attractor solutions can help in improving the solution towards the fully non-linear regime. We do not expect our equation to be valid in the truly non-linear regime where intrinsic dissipation effects lead to virialized structures. We do expect to be able to understand its limitations in a quantitative way and consequently the resulting  limits on the precision of its solution in the quasi-linear regime.

\section*{Acknowledgments}
\noindent
We thank Massimo Pietroni, Roman Scoccimarro, Filippo Vernizzi and Yair Zarmi for discussions.
The research of RB was supported by The Israel Science Foundation grant no. 239/10.

\bibliographystyle{JHEP}

\end{document}